\documentclass[aps,reprint,twocolumn,letterpaper,showpacs,preprintnumbers,amsmath,amssymb]{revtex4}
\usepackage{epsfig,amssymb}
\usepackage{graphicx}
\usepackage{dcolumn}
\usepackage{bm}
\usepackage{dcolumn}
\usepackage{multirow}
\begin{document}
\title{A joint time-dependent density-functional theory for excited
states of electronic systems in solution}
\author{Johannes~Lischner}
\affiliation{Department of Physics, University of California, Berkeley, California 94720, USA, and Materials Sciences Division, Lawrence Berkeley National Laboratory, Berkeley 94720, USA.}
\author{T.A.~Arias}
\affiliation{Laboratory of Atomic and Solid State Physics, Cornell University, Ithaca, New
York 14853, USA}

\begin{abstract}
We present a novel joint time-dependent density-functional theory for
the description of solute-solvent systems in time-dependent external
potentials. Starting with the exact quantum-mechanical action
functional for both electrons and nuclei, we systematically eliminate
solvent degrees of freedom and thus arrive at coarse-grained action
functionals which retain the highly accurate \emph{ab initio}
description for the solute and are, in principle, exact.  This
procedure allows us to examine approximations underlying popular
embedding theories for excited states. Finally, we introduce a novel
approximate action functional for the solute-water system and compute
the solvato-chromic shift of the lowest singlet excited state of
formaldehyde in aqueous solution, which is in good agreement with
experimental findings.
\end{abstract}

\maketitle

\section{Introduction}
Electronic excited states are important in many areas of physics,
chemistry and biology. They are probed in spectroscopic experiments,
such as absorption, reflectivity or photoluminescence measurements.
In addition, they are important in many technical applications, such
as photovoltaics \cite{ScholesNature,LouiePPV}, laser technology
\cite{Heeger2} or light-emitting diodes\cite{Holmes,Salaneck}.
However, in most situations the excited system is not in isolation,
but in contact with an environment. For example, dyes in Gr\"atzel
cells \cite{Selloni,Graetzel} are in contact with an electrolyte and
spectra of DNA molecules are typically obtained in an aqueous
solution \cite{Rubio,Improta}.

Various methods for the theoretical modelling of electronic excited
states have been developed. For extended systems, such as periodic
solids or surfaces, excitation energies are typically extracted from
the single-particle and two-particle Green's functions by solving the
quasiparticle \cite{LouieGW} and the Bethe-Salpeter equation
\cite{LouieBSE} in the GW approximation. Quantum chemistry methods,
such as configuration interaction \cite{Cramer} or coupled cluster
theory \cite{Bartlett}, yield highly accurate excitation energies for
atoms and small molecules. In contrast to the aforementioned methods,
which scale unfavorably with the system size, in recent years
time-dependent density-functional theory\cite{RungeGross,Casida}
emerged as an economical yet accurate theory for larger molecules and
clusters \cite{Martin,AhlrichC60,Selloni}.

However, despite its good scaling properties, the application of
time-dependent density-functional theory to electronic systems, which
are not in isolation, but in contact with an environment, remains
numerically challenging. To capture solvent effects on excited states,
a number of embedding approaches have been developed
\cite{Barone,Mikkelsen,Truhlar,Coutinho,Ruud}. These methods either
model the solvent atomistically, for example by using the classical
molecular dynamics technique\cite{Mikkelsen,Coutinho}, or via a
continuum approach\cite{Barone,Ruud,Truhlar}. Due to its simplicity,
the latter approach has enjoyed great popularity. In particular, many
calculations employed time-dependent density-functional theory in
conjunction with the ``polarizable continuum model''
\cite{Barone,Selloni,Adamo,Graetzel}, where the molecule is placed
inside a cavity in a linear dielectric medium. The solute-solvent
interactions are then separated into equilibrium and non-equilibrium
contributions accounting for the fact that the electronic excitations
on the molecule are screened by the high-frequency dielectric
constant, which in many systems is much smaller than the static
dielectric constant.

Despite the success of these continuum models, it is important to
recall that their construction is purely phenomenological. To improve
upon these theories and understand their limitations, it is necessary
to understand their origin \emph{from first principles}. In this
paper, we derive different levels of continuum embedding theory
starting from the \emph{exact} quantum-mechanical action functional
for the full solute-solvent system (Sections~\ref{sec:JTDDFT}). Next,
as a demonstration that the approach can lead to practical
calculations, we introduce a novel time-dependent continuum model
(Section~\ref{sec:water-solute}), which takes into account the
deviation from bulk behavior of the solvent response in the first
solvation shells and also retains the full frequency dependence of the
dielectric response, and then apply this functional to the excitations
of a formaldehyde molecule in aqueous solution
(Section~\ref{sec:formaldehyde}). Finally, in
Section~\ref{sec:summary} we discuss our conclusions and describe
possible future developments and applications.

\section{Joint time-dependent density-functional theory}
\label{sec:JTDDFT}

In this section, we consider a physical system composed of an explicit
subsystem ({\em solute}), in contact with an environment ({\em
solvent}). Both solute and solvent consist of electrons and nuclei
of various species. Examples of such solute-solvent systems are a
molecule dissolved in water or a defect in a host crystal.
To investigate the behavior of a solute-solvent system in a
time-dependent external potential, we employ time-dependent
density-functional theory \cite{RungeGross}.  Here, we begin with the
straightforward generalization of the standard, fully causal
expression for the action of a single-component system at zero
temperature \cite{vanLeeuwen} to multi-component systems at finite
temperature.  This generalization is similar to the theory of Li and
Tong \cite{LiTong}.  Those authors, however, worked only at zero
temperature and employed the Frenkel-Dirac action functional, which
violates the causality requirement of van Leeuwen \cite{vanLeeuwen}.
Generalizing the approach of van Leeuwen to the present situation
yields for the action functional $S$ of the full solute-solvent
system,
\begin{align}
S[n,\{N_\alpha\}] =& A[n,\{N_\alpha\}] - \int_\Omega d\bm{r}\int_C d\tau
n(\bm{r},\tau)v(\bm{r},\tau)  \nonumber\\
&- \sum_\alpha \int_\Omega d\bm{r} \int_C d\tau
N_\alpha(\bm{r},\tau)V_{\alpha}(\bm{r},\tau),
\label{Sfull}
\end{align}
where $n(\bm{r})$ and $N_{\alpha}(\bm{r})$ denote the density of
electrons and the various nuclear species, while $v(\bm{r})$ and
$V_{\alpha}(\bm{r})$ denote the respective external potentials.  Also,
$\Omega$ denotes an open volume as required when working in the grand
canonical ensemble and $\tau$ denotes the Keldysh time \cite{Keldysh},
which is defined on the contour $C$ ranging from $0$ to $\infty$ just
above the real time axis and then back from $\infty$ to $0$ just under the
real time axis, and finally from $0$ to $-i\beta$ on the imaginary
time axis, with $\beta=1/(k_BT)$ being the inverse thermal
energy. Finally, in \eqref{Sfull}, the intrinsic action $A$ is the
Legendre transform, with respect to the potentials $v$ and $V_\alpha$
\cite{vanLeeuwen}, of
\begin{align}
&\tilde{A}[v,\{V_{\alpha}\}]=  \nonumber \\
&i\log {\rm Tr} \left\{
\exp\left( \beta
\left[ \mu_{\rm el}\hat{N}_{\rm el}+\sum_{\alpha}\mu_{\alpha}\hat{N}_\alpha \right]
\right) \hat{U}(-i\beta;0) \right\},
\label{eq:A}
\end{align}
where $\mu_{\rm el}$ and $\mu_\alpha$ are the chemical potentials of
the electrons and nuclei, while $\hat{N}_{\rm el}$ and $\hat{N}_
\alpha$ denote the respective particle number operators. In the above,
$\hat{U}=T_C \exp(-i\int d\tau \hat{H}(\tau))$ denotes the
quantum-mechanical evolution operator\cite{vanLeeuwen} with $T_C$
being the Keldysh time-ordering operator and $\hat{H}(\tau)$ denotes
the standard many-body Hamiltonian for the electrons and nuclei of the
solute-solvent system.  

We note that, relatively recently, Butriy {\em et al.} have also
considered multicomponent time-dependent density-functional theory
\cite{Gross}, but chose as the variational variables the electron
density in body-fixed coordinates and the diagonal of the nuclear
N-body density matrix, whereas we here employ as the variables the
much more tractable densities $n(\bm{r},\tau)$ and
$N_\alpha(\bm{r},\tau)$.  Butriy {\em et al.} employed their formalism
to study correlated electron-nuclear excitations in isolated
molecules. By contrast, we here are interested in electronic
excitations of the system while treating the solute nuclei in the
Born-Oppenheimer approximation, holding them fixed in place, so that
they present a simple {\em fixed external potential} in which the
electrons and solvent nuclei evolve. With the solute nuclear
coordinates fixed, there is no need to work in body-fixed coordinates
and a simple density description is sufficient for us to extract the
density fluctuations of interest.

We do, however, find it mathematically convenient {\em as a matter of
  bookkeeping} to treat the solute and environment nuclear densities
on an equal footing for as long as possible; therefore, we treat the
solute nuclear densities as time-dependent in our derivation and only
fix the locations of the solute nuclei in the final
step. Consequently, the index $\alpha$ in \eqref{Sfull} and
\eqref{eq:A} ranges over all nuclear species in both the solvent and
the solute.

Because of the many environment degrees of freedom, finding the
time-dependent densities which make the action in \eqref{Sfull}
stationary is numerically challenging. Moreover, the explicit details
of the density fields describing the solvent are often irrelevant,
because one is typically interested in properties of the solute. We
therefore seek a fundamental description which treats the solute
explicitly and the solvent either at a simplified level or implicitly.
Petrosyan and coworkers \cite{Petrosyan1, Petrosyan2} have developed
just such a rigorous ``joint'' density-functional theory for the
static, equilibrium case. Specifically, Petrosyan {\em et al.}  first
minimize the full solute-solvent free-energy functional over the
solvent electron density to arrive at a free-energy functional in
terms of the solute electron and the solvent nuclear densities. The
resulting theory treats both solute and solvent explicitly, but the
solvent at a more coarse-grained level.  Ultimately, for specific
solute-solvent systems the coarse-grained free-energy functional is
minimized over all solute electron and solvent nuclear densities to
obtain the free energy of the overall system and its equilibrium
properties. Petrosyan {\em et al.}  developed accurate and numerically
tractable approximations to the coarse-grained functional and employed
them to study surfaces and small molecules in aqueous solution with
encouraging results\cite{Petrosyan1,Petrosyan2}.

To generalize the theory of Petrosyan and coworkers to the present
non-equilibrium context, we split the total electron density in
\eqref{Sfull} into solute ($n_s$) and environment ($n_e$)
contributions,
$n(\bm{r},\tau)=n_s(\bm{r},\tau)+n_e(\bm{r},\tau)$. Fundamentally, a
rigorous partitioning of electrons into solute and environment
electrons is, of course, impossible because of the their
quantum-mechanical indistinguishability. Nonetheless, making $S$
stationary with respect to all physically allowed environment electron
densities and subsequently with respect to all physically allowed
solute electron densities is guaranteed to recover the correct total
electron density. There are, of course, many ways to express the total
electron density as a sum of two subsystem densities.  Thus, instead
of a unique solution, there exists, in principle, a vast degenerate
set of solutions in joint time-dependent density-functional theory
consisting of all solute and environment electron densities which sum
up to the correct total electron density. In practice, however, we
find that practical approximations break this degeneracy and pick out
a sensible particular solution.  This is reminiscent of the
equilibrium case, where Petrosyan \emph{et al.} \cite{Petrosyan2}
observed that the use of molecular pseudopotentials
\cite{Joannopoulos} leads to sensible non-degenerate solutions.

\subsection{Explicit solvent functionals}
Making the action stationary with respect to the environment electron
density while holding the solute electron and environment nuclear
densities fixed, we obtain the coarse-grained explicit-solvent
functional $S_{\rm ex}$,
\begin{widetext}
\begin{align}
S_{\rm ex}[n_s,\{N_\alpha\}]=& {\rm stat}_{n_e}
\left\{ \vphantom{\int} A[n_s+n_e,\{N_\alpha\}]
-\int_\Omega d\bm{r} \int_C d\tau \left[ n_e(\bm{r},\tau) v(\bm{r},\tau)
+\sum_{\alpha}  N_\alpha(\bm{r},\tau)V_\alpha(\bm{r},\tau) \right] \right\}
\nonumber\\
&-\int_\Omega d\bm{r} \int_C d\tau n_s(\bm{r},\tau)v(\bm{r},\tau)
\nonumber\\
&\equiv {\cal A}^{(v,\{V_\alpha\})}[n_s,\{N_\alpha\}]
-\int_\Omega d\bm{r} \int_C d\tau n_s(\bm{r},\tau)v(\bm{r},\tau),
\label{SCG1}
\end{align}
\end{widetext}
where ${\rm stat}_{n_e}$ indicates that the expression in curly
brackets is made stationary with respect to variations of $n_e$ and
the superscript of ${\cal A}$ makes explicit the additional dependence
of this functional on the external potentials.  Note that we include
the coupling term for the nuclear densities ($\sum N_\alpha V_\alpha$)
in ${\cal A}$: this partitioning is not functionally necessary,
because the coupling term does not depend on $n_e$ and therefore
maintains its simple form.  This choice, however, ensures that ${\cal
  A}$ describes a neutral system: Maintaining charge neutrality is
important both formally to ensure the existence of the thermodynamic
limit \cite{LiebLebowitz} and also practically to mitigate the need to
capture long-range couplings within an approximate functional. We
stress that our partitioning does not fundamentally complicate the
functional dependence of the new functional ${\cal A}$ because the
coupling to $V_\alpha$ will retain its simple form in terms of
$N_\alpha$.

To find practical approximations, we partition $\cal{A}$ into various
physically meaningful contributions according to
\begin{widetext}
\begin{align}
{\cal A}= &
A_{\rm el,s}[n_s] + A_{\rm nuc,s}[\{N_{s,\alpha}\}]
-A^{(\{ V_{s,\alpha}\})}_{\rm nuc/ext,s}[\{ N_{s,\alpha} \}] -
A_{\rm nuc/el,s}[n_s,\{N_{s,\alpha}\}] \nonumber \\
& + A_e[\{ N_{e,\alpha} \}]
- \Delta {\cal A}^{(v,\{ V_{e,\alpha}\})}_{\rm ext,e}[\{ N_{e,\alpha} \}]
- \Delta {\cal A}^{(v,\{ V_{e,\alpha}\},\{V_{s,\alpha}\})}_{s,e}[n_s,\{ N_{s,\alpha} \},\{ N_{e,\alpha} \}],
\label{ACG1}
\end{align}
\end{widetext}
where the first four terms describe the solute: The first term,
$A_{\rm el,s}$, denotes the intrinsic action of the solute electrons
and is typically \cite{vanLeeuwen} written as
\begin{equation}
A_{\rm el,s} = A_{KS}-A_H-A_{XC},
\label{AKS}
\end{equation}
where $A_{KS}$ denotes the action of non-interacting electrons,
$A_H=1/2 \int d\bm{r} \int d\bm{r}' \int d\tau
n_s(\bm{r},\tau)n_s(\bm{r}',\tau’)/|\bm{r}-\bm{r}'|$ is the Hartree
contribution and $A_{XC}$ the exchange-correlation term. The
second and third terms in \eqref{ACG1}, respectively, are the
intrinsic action of the solute nuclei (with densities $N_{s,\alpha}$)
and their coupling to the external potentials. The fourth term
captures the interaction between solute electrons and solute
nuclei. In our actual calculations, we hold the solute nuclei fixed in
space, finding $A_{\rm nuc/el,s}=\int d\bm{r} \int d\tau
n_s(\bm{r},\tau)v_{st}(\bm{r})$, with $v_{st}$ being the static
potential created by the solute nuclei. Under these conditions, the
second and third term in \eqref{ACG1} become independent of all
time-dependent {\em degrees of freedom} (though, not of time-dependent
potentials) and can be dropped for the purpose of the variational
calculations. However, for later convenience we retain $A_{\rm
  nuc,s}=\int d\tau \sum_{I<J} Z_I Z_J/|\bm{R}_I-\bm{R}_J|$ with
$\bm{R}_I$ and $Z_I$ denoting the positions and charges of the solute
nuclei.

The fifth term in \eqref{ACG1} describes the isolated neutral
environment in terms of its nuclear densities,
$N_{e,\alpha}$. Accurate approximations for this action functional are
less well known than for electrons. However, there has been much
progress recently in the construction of such action functionals for
classical liquids \cite{Chan,Archer}, which constitute an important
and technologically relevant class of solvents.

Finally, having identified all interactions between \emph{charged}
species, we expect the remaining two contributions in \eqref{ACG1} to
be relatively small.  The sixth term, $\Delta {\cal A}_{\rm ext,e}$,
describes the interaction between the \emph{neutral} solvent and the
external potentials.  The final, seventh term, $\Delta {\cal
  A}_{s,e}$, maintains the full functional dependence and captures
\emph{by definition} all remaining interactions. We find that the
coupling between the \emph{neutral} solute and the \emph{neutral}
solvent constitutes the most important contribution to this term.

To find approximations to $\Delta {\cal A}_{\rm ext,e}$, we observe
that the best form for this term depends on the physical system under
consideration, because the solvent electrons screen the bare nuclear
charges in \emph{qualitatively different} ways depending on the
physical nature of the solvent. If the solvent consists of ions of
charge $\bar{Z}_\alpha$, the corresponding action would be
$\sum_{\alpha}\int d\tau \int d\bm{r} [N_{e,\alpha} V_\alpha +
  (Z_\alpha-\bar{Z}_\alpha)N_{e,\alpha} v]$ with $Z_\alpha$ being the
true charge of the nucleus.  If, however, the solvent is composed of
neutral polar molecules, where each ``effective'' nucleus carries a
partial charge $q_\alpha$ and all partial charges in a molecule add up
to zero, one should replace $\bar{Z}_\alpha$ in the above expression
by $q_\alpha$. Finally, if the solvent consists of apolar molecules or
neutral atoms, we can approximate the coupling by $\sum_\alpha\int
d\tau \int d\bm{r} \pi_\alpha N_{e,\alpha}|\nabla v|^2$ with
$\pi_\alpha$ being the polarizability of the ``effective'' nucleus
$\alpha$.

The final term, $\Delta{\cal A}_{s,e}$, has the full functional
dependence and thus can capture all remaining interactions.  In
typical density-functional theory fashion, because we have separated
out by various approximations all other possible interactions and
ensured that this term represents a charge-neutral interaction, we
expect this term to be relatively small with mild functional
dependencies and thus amenable to simple approximations. Accordingly,
we expand $\Delta {\cal A}_{s,e}$ as a Taylor series in the various
densities, keeping only the lowest-order coupling terms,
\begin{widetext}
\begin{align}
\Delta{\cal A}_{s,e} =
\int_\Omega d\bm{r} \int_\Omega d\bm{r}' \int_C d\tau \int_C d\tau'
\sum_\alpha N_{e,\alpha}(\bm{r}',\tau') \left[ \vphantom{\int}
w_{\alpha}(\bm{r},\bm{r}',\tau,\tau') n_s(\bm{r},\tau)
+\sum_\beta w_{\alpha\beta}(\bm{r},\bm{r}',\tau,\tau') N_{s,\beta}(\bm{r}',\tau')
\right],
\label{DeltaA}
\end{align}
\end{widetext}
where $w_{\alpha}(\bm{r},\bm{r}',\tau,\tau')=\delta^2 \Delta {\cal
  A}_{s,e}/\delta n_s(\bm{r},\tau)\delta N_{e,\alpha}(\bm{r}',\tau')$
and $w_{\alpha\beta}(\bm{r},\bm{r}',\tau,\tau')=\delta^2 \Delta {\cal
  A}_{s,e}/\delta N_{e,\alpha}(\bm{r},\tau)\delta
N_{s,\beta}(\bm{r}',\tau')$ denote effective time-dependent
interaction potentials between solvent nuclei and solute electrons or
solute nuclei. Note that, in principle, the Taylor series contains
various other coupling terms: for example, a term quadratic in the
solute electron density can occur. However, such a term only
renormalizes the Hartree contribution in $\cal{A_{\rm el,s}}$ and is
therefore neglected in \eqref{DeltaA}.

Table~\ref{AllTerms} summarizes all of the above considerations,
listing all of the various contributions to $S_{\rm ex}$ and the terms
which capture them.

\begin{table}
\begin{center}
  \setlength{\doublerulesep}{1\doublerulesep}
  \setlength{\tabcolsep}{3\tabcolsep}
  \caption{The Table shows the various contributions to $S_{\rm ex}$
    due to interactions between solute nuclei ($N_s$), solute
    electrons ($n_s$), solvent nuclei ($N_e$), solvent electrons
    ($n_e$) and the external potentials ($v$ and $V$). Note that in
    $S_{ex}$ the solvent electrons are treated implicitly.\\}
  \begin{ruledtabular}
    \begin{tabular}{c c}
      contribution & contained in \\
      \hline
       & \\
      $N_s$/$N_s$    & $A_{\rm nuc,s}$ \\
      $N_s$/$n_s$    & $A_{\rm nuc/el,s}$\\
      $N_s$/$N_e$    & $\Delta {\cal A}_{s,e}$ \\
      $N_s$/$n_e$    & $\Delta {\cal A}_{s,e}$ \\
      $N_S$/$V$      & $A_{\rm nuc/ext,s}$ \\
      $n_s$/$n_s$    & $A_{\rm el,s}$ \\
      $n_s$/$N_e$    & $\Delta {\cal A}_{s,e}$ \\
      $n_s$/$n_e$    & $\Delta {\cal A}_{s,e}$ \\
      $n_s$/$v$      & last term in \eqref{SCG1} \\
      $N_e$/$N_e$    & $A_e$ \\
      $N_e$/$n_e$    & $A_e$ \\
      $N_e$/$V$      & $\Delta {\cal A}_{\rm ext,e}$ \\
      $n_e$/$n_e$    & $A_e$ \\
      $n_e$/$v$      & $\Delta {\cal A}_{\rm ext,e}$ \\
    \end{tabular}
  \end{ruledtabular}
  \label{AllTerms}
\end{center}
\end{table}

\subsection{Implicit solvent functionals}
\label{sec:implicit}
\subsubsection{General considerations}

Rather than follow the above route of dealing explicitly with the
solvent nuclei, for this initial work, we take a simpler tack that
allows us to make contact with standard continuum solvent models. For
this purpose, we eliminate the environment nuclei from $S_{\rm ex}$
and introduce a new action functional $S_{\rm im}$ which depends on
the solute densities \emph{only} and treats the solvent
\emph{implicitly} as follows,
\begin{align}
&S_{\rm im}[n_s,\{N_{s,\alpha}\}] = {\rm stat}_{\{N_{e,\alpha}\}}
{\cal A}^{(v,\{V_{\alpha}\})}[n_s,\{N_{s,\alpha}\},\{N_{e,\alpha}\}]
\nonumber\\
& \qquad \qquad \qquad \qquad
-\int_\Omega d\bm{r} \int_C d\tau n_s(\bm{r},\tau)v(\bm{r},\tau)
\nonumber \\
&\equiv {\cal G}^{(v,\{V_{\alpha}\})}[n_s,\{N_{s,\alpha}\}]
-\int_\Omega d\bm{r} \int_C d\tau n_s(\bm{r},\tau)v(\bm{r},\tau).
\label{SCG2}
\end{align}
Again, we partition ${\cal G}$ into meaningful contributions according
to
\begin{align}
{\cal G}&=
A_{\rm el,s}[n_s] +A_{\rm nuc,s}[\{N_{s,\alpha}\}] 
-A^{(\{V_{s,\alpha}\})}_{\rm nuc/ext,s}[\{N_{s,\alpha}\}]
\nonumber \\
&-A_{\rm nuc/el,s}[n_s,\{N_{s,\alpha}\}]
-\Delta{\cal G}^{(v,\{V_{e,\alpha}\})}[n_s,\{N_{s,\alpha}\}]
\label{G}
\end{align}
with $\Delta{\cal G}=-{\rm stat}_{\{ N_{e,\alpha}\}}[A_e-\Delta {\cal
    A}_{\rm ext,e}-\Delta {\cal A}_{s,e}]$ and $\cal{A_{\rm el,s}}$ is
given by \eqref{AKS}.  Note that $\Delta{\cal G}$ depends on the
solute densities of electrons and nuclei, but also on the
\emph{time-dependent external potential}. To understand the
consequences of the additional functional dependency, we now
investigate the linear response behavior of $S_{\rm im}$ in greater
detail (assuming fixed solute nuclei).  

The time-dependent solute electron density corresponding to $v$ makes
$S_{im}$ stationary, $\delta S_{im}=0$, which implies $\delta {\cal
  G}/\delta n_s=v$. Using \eqref{G} we thus have (assuming fixed
solute nuclei)
\begin{align}
v(\bm{r},\tau)=&v_{KS}(\bm{r},\tau)-v_H(\bm{r},\tau)-
v_{XC}(\bm{r},\tau) \nonumber\\
&- v_{st}(\bm{r}) - v^{(v,\{V_{e,\alpha}\})}_{e}(\bm{r},\tau),
\label{v_KS}
\end{align}
where $v_{KS}=\delta A_{KS}/\delta n_s$ denotes the Kohn-Sham
potential, $v_H=\delta A_H/\delta n_s$, $v_{XC}=\delta A_{XC}/\delta
n_s$ and $v_{st}$ is the static potential due to the solute nuclei.
Also, $v_{e}=\delta \Delta {\cal G}/\delta n_s$ denotes the additional
potential due to the presence of the environment. Note that
$\Delta{\cal G}$ and therefore also $v_{e}$ depend both on the solute
density $n_{s}$ \emph{and} the external potential \emph{separately}.

A small change $\delta v$ in the external potential causes a change
$\delta n_{s}$ in the solute electron density.  In the linear response
regime, these quantities are related via the response function $\chi$
\begin{equation}
\delta n_{s}(\bm{r},\tau)=\int_{\Omega} d\bm{r}' \int_C d\tau'  
\chi(\bm{r},\bm{r}',\tau,\tau') \delta v(\bm{r}',\tau').
\label{chi}
\end{equation}
To compute $\chi$, which is the observable in spectroscopic
experiments on the solute-solvent system, we first determine the
change in the Kohn-Sham potential $\delta v_{KS}$ corresponding to
$\delta v$. (Strictly speaking, only the \emph{total} response
function $\chi_{\rm tot}=\delta n/\delta v=\delta n_s/\delta v +
\delta n_e/\delta v$ is measured. However, for solute-solvent
systems where the response of the solute occurs in a different
frequency range than the response of the solvent one can determine
experimentally the solute response function $\chi$. This is the case
for the lowest singlet excitation of formaldehyde in water, which we
study in the Section \ref{sec:formaldehyde}.) Using \eqref{v_KS}, we
find
\begin{align}
\delta v_{KS}(\bm{r},\tau)&=\delta v(\bm{r},\tau) + \int_\Omega d\bm{r}'
\frac{\delta n_s(\bm{r}',\tau)}{|\bm{r}-\bm{r}'|}  \nonumber \\
&+\int_\Omega d\bm{r}' \int_C d\tau'
f_{XC}(\bm{r},\bm{r}',\tau,\tau')\delta n_s(\bm{r}',\tau')  
\nonumber \\  
&+\int_\Omega d\bm{r}' \int_C d\tau'
f^A_e(\bm{r},\bm{r}',\tau,\tau')\delta n_s(\bm{r}',\tau') \nonumber \\
&+\int_\Omega d\bm{r}' \int_C d\tau'  
f^B_e(\bm{r},\bm{r}',\tau,\tau')\delta v(\bm{r}',\tau') \nonumber \\
&+\sum_\alpha \int_\Omega d\bm{r}' \int_C d\tau'  
f^C_{e,\alpha}(\bm{r},\bm{r}',\tau,\tau')\delta V_{e,\alpha}(\bm{r}',\tau'),
\label{dv_KS}
\end{align}
where $f^A_e=\delta v_e/\delta n_s$, $f^B_e=\delta v_e/\delta v$ and
$f^C_{e,\alpha}=\delta v_e/\delta V_{e,\alpha}$ denote
additional contributions to $\delta v_{KS}$ caused by the
environment. In an actual experiment, where the whole solute-solvent
system is probed, for example by an electromagnetic wave, we expect
$\delta V_{e,\alpha}$ to be related to $\delta v$. In this case, we
can express the last term in \eqref{dv_KS} as $f^D_e \delta v$ with 
$f^D_e = \sum_\alpha f^C_{e,\alpha} \delta V_{e,\alpha}/\delta v$.

The change in the Kohn-Sham potential is related to $\delta n_s$ via
\begin{equation}
\delta n_{s}(\bm{r},\tau)=\int_{\Omega} d\bm{r}' \int_C d\tau'  
\chi_{KS}(\bm{r},\bm{r}',\tau,\tau') \delta v_{KS}(\bm{r}',\tau'),
\label{chi_KS}
\end{equation}
where $\chi_{KS}$ denotes the response function of non-interacting
electrons. Combining \eqref{chi_KS}, \eqref{dv_KS} and \eqref{chi} and
adopting a matrix formulation for the space and Keldysh-time variables
then yields
\begin{align}
\chi^{-1}  =[1+f^B_e + f^D_e]^{-1} \left\{
\chi^{-1}_{KS}- [ K+f_{XC}+f^A_e ] \right\},
\label{chiSolv}
\end{align}
where $K$ denotes the matrix corresponding to the Coulomb interaction
$K(\bm{r},\bm{r}',\tau,\tau')=\delta(\tau,\tau')/|\bm{r}-\bm{r}'|$
with $\delta(\tau,\tau')$ denoting the Delta-function on the Keldysh
contour.

Compared to the familiar equation for $\chi$ \emph{without solvent},
given by $\chi^{-1}=\chi^{-1}_{KS}-[K+f_{XC}]$, we find that
\eqref{chiSolv} contains \emph{three} extra terms due to the presence
of the solvent: $f^A_e$ describes the change of the solvent potential
due to a change in the solute electron density, while $f^B_e$ and
$f^D_e$ describe changes induced by a variation in the external
potential.
Without justification, ``polarizable continuum model'' approaches
typically \cite{Barone} approximate the potential due to the solvent
as a functional of the solute density \emph{only}, which means they
only include $f^A_e$ and \emph{neglect} $f^B_e$ and $f^D_e$. This
insight into the assumptions underlying popular embedding approaches
underscores the value of following the density-functional approach
rigorously, so as to identify all potentially relevant functional
dependencies.

\subsubsection{Practical approximations}

To develop practical approximations, we separate $\Delta{\cal G}$ into
a contribution $\Delta{\cal G}_{s,e}[n_s,\{N_{s,\alpha}\}]$, which
describes the interaction among solute particles \emph{mediated by the
environment}, and a remainder $\Delta{\cal
G}^{(v,\{V_{e,\alpha}\})}_{\rm ext}[n_s,\{N_{s,\alpha}\}]$. Taylor expanding
$\Delta{\cal G}_{s,e}$ yields
\begin{widetext}
\begin{align}
\Delta{\cal G}_{s,e}= & \int_\Omega d\bm{r} \int_\Omega d\bm{r'} \int_C d\tau
\int_C d\tau'
\left[ \frac{1}{2} n_s(\bm{r},\tau) W(\bm{r},\bm{r'},\tau,\tau') n_s(\bm{r'},\tau')
\right. \nonumber \\
&\left. + \sum_\alpha N_{s,\alpha}(\bm{r},\tau)
\left( W_\alpha(\bm{r},\bm{r'},\tau,\tau') n_s(\bm{r'},\tau')
+ \frac{1}{2}\sum_\beta W_{\alpha\beta}(\bm{r},\bm{r'},\tau,\tau')
N_{s,\beta}(\bm{r'},\tau')
\right) \right],
\label{Gse}
\end{align}
\end{widetext}
where $W(\bm{r},\bm{r'},\tau,\tau')$,
$W_\alpha(\bm{r},\bm{r'},\tau,\tau')$ and
$W_{\alpha\beta}(\bm{r},\bm{r'},\tau,\tau')$ denote \emph{effective
interaction potentials} between the various solute particles. In the
next section, we approximate these interaction potentials by screened
Coulomb interactions, which results in a simple, yet accurate joint
density-functional theory for solute-water systems.

Approximating $\Delta{\cal G}_{\rm ext}$ is more difficult: a possible
route to finding explicit functionals is to express the environment
nuclear densities in terms of solute densities according to
$N_{e,\alpha}(\bm{r})=g_{\alpha}[n_s,\{N_{s,\beta}\}](\bm{r})$ and
insert this relation into the various forms for $\Delta{\cal A}_{\rm
  ext,e}$ discussed in the last section. We expect that $g_\alpha$ has
a similar form as the dielectric function employed in the next section
[Equation~\eqref{epsilon}], where we employ a local ansatz to describe
the crossover from bulk screening to vacuum.

However, for our first implementation of joint time-dependent
density-functional theory for solute-water systems presented in the
next section we neglect $\Delta{\cal G}_{\rm ext}$. We expect,
however, that the additional solvent response due to this term can be
included by ``renormalizing'' the dielectric function describing the
environment (see Section \ref{sec:water-solute}). Future work should
explore the consequences and importance of this term.

\section{An implicit action functional for the water-solute system}
\label{sec:water-solute}
To allow us to explore and test the potential of the above ideas in an
actual application, in this section, we introduce a relatively simple,
approximate joint time-dependent density-functional for the
solute-water system. In particular, we assume that all solvent effects
can be described via a position \emph{and} frequency-dependent
local dielectric function, which depends on the electronic structure of the
solute. Inclusion of the spatial dependence of screening effects is
crucial, because the dielectric response of water in the first
solvation shells differs notably from the bulk response. Also, in
contrast to ``polarizable continuum model'' approaches
\cite{Barone,MennucciCammi}, where a particular value for the
high-frequency dielectric constant is chosen, we employ the full
frequency-dependent dielectric function.

Specifically, the assumption of dielectric screening implies that all
effective interactions introduced in \eqref{Gse} are proportional to
the screened interaction $\tilde{W}$ between two unit charges and only
rescaled by the charges of the interacting species. In particular, we
approximate $W=\tilde{W}$, $W_{\alpha}=-Z_{\alpha}\tilde{W}$ and
$W_{\alpha\beta}=Z_\alpha Z_\beta \tilde{W}$. The resulting action
functional for the solute-water system is then given by
\begin{widetext}
\begin{align}
{\cal G} = A_{KS}-A_{XC}- \Delta V_{ps}
- \frac{1}{2}\int_\Omega d\bm{r} \int_\Omega d\bm{r'}
\int_C d\tau \int_C d\tau'
\rho_s(\bm{r},\tau) \tilde{K}(\bm{r},\bm{r'},\tau,\tau') \rho_s(\bm{r'},\tau'),
\label{Sapprox}
\end{align}
\end{widetext}
with $\rho_s(\bm{r},\tau)=-n_s(\bm{r},\tau)+\sum_I
Z_I\delta(\bm{r}-\bm{R}_I)$ denoting the solute charge density and
$\tilde{K}=K+\tilde{W}$, where $K$ is the bare Coulomb interaction
defined as above. Also, $\Delta V_{ps}$ reflects the fact that, in
practical calculations, we employ the {\em pseudopotential
  approximation} \cite{TeterArias}, in which the nuclei are replaced
with ionic cores of charge $Z_I$, whose potentials at large distances
(when not screened by the environment) go as $Z_I/|\bm{r}-\bm{R}_I|$
but which differ from this by a localized function $\Delta
V_{ps}(\bm{r}-\bm{R}_I)$ within a small ``core radius'' that
represents a distance much smaller than where we would expect
screening from the environment to occur.  Within our framework, the
long-range parts enter through the solution of \eqref{ScreenedPoisson}
and thus are properly screened, and the short-range parts contained in
$\Delta V_{ps}$ enter directly as they require no such screening.

The screened potential corresponding to a \emph{physical} charge
density $\rho_s(\bm{r},t)$, which is equal on both vertical branches
of the Keldysh contour, is given by $\tilde{\phi}_s \equiv
\tilde{K}^R\rho_s$, where $\tilde{K}^R$ denotes the retarded
interaction \cite{vanLeeuwen}. In actual calculations, we obtain
$\tilde{\phi}_s$ by solving the screened Poisson equation
\begin{align}
\nabla\cdot \epsilon(\bm{r},\omega) \nabla \tilde{\phi}(\bm{r},\omega)=
-4\pi\rho_s(\bm{r},\omega).
\label{ScreenedPoisson}
\end{align}
All information about the environment is contained in the dielectric
function $\epsilon(\bm{r},\omega)$. In principle, both the ionic and
the electronic degrees of freedom of the solvent contribute to the
dielectric response. We demonstrate below that for the frequencies of
interest, we can safely ignore the motion of the ions and only deal
with the electronic response corresponding to a fixed nuclear solvent
density. We make the natural assumptions that the system is in
equilibrium before the excitation and that the equilibrium nuclear
solvent density is determined \emph{locally} by the equilibrium solute
electron density $n_0(\bm{r})$. This suggests the following local ansatz
for the dielectric function,
\begin{equation}
\epsilon(\bm{r},\omega)=\epsilon(n_0(\bm{r}),\omega).
\end{equation}
This ansatz is physically reasonable in that it interpolates smoothly
between the dielectric response of vacuum and the bulk liquid and thus
avoids the need to specify a cavity shape.
If we further assume that the frequency dependence of
$\epsilon(n_0(\bm{r}),\omega)$ enters only through the frequency
dependence of the bulk dielectric function $\epsilon_b(\omega)$, we
can generalize the form employed by Petrosyan and co-workers
\cite{Petrosyan1} to
\begin{equation}
\epsilon(\bm{r},\omega)=1+\frac{\epsilon_b(\omega)-1}{2}
{\rm erfc}\left( \frac{{\rm log}(n_0(\bm{r})/n_c)}{\sqrt{2}\sigma} \right),
\label{epsilon}
\end{equation}
where the parameters $n_c$ and $\sigma$ determine the location and
width, respectively, of the crossover from the vacuum to the bulk
liquid dielectric response. Petrosyan {\em et al.} \cite{Petrosyan1}
determined the numerical values $n_c=4.73\times 10^{-3}\AA^{-3}$ and
$\sigma=0.6$ for these parameters by fitting solvation energies of
small molecules obtained by their equilibrium joint density-functional
theory to experimental data.  We choose to work with these values as
well.

To complete the theory, we need an expression for the
frequency-dependent bulk dielectric constant $\epsilon_b(\omega)$. At
frequencies corresponding to electronic excitations, we may ignore the
complicated low-frequency dielectric response of water and employ a
model which describes the high-frequency range reliably. For this, we
use the Clausius-Mossotti form \cite{AshcroftMermin}
\begin{equation}
\frac{ \epsilon_b(\omega) -1}{\epsilon_b(\omega)+2}=
\frac{ 4\pi }{3} n_b \bar{\alpha}(\omega),
\label{eps_bulk}
\end{equation}
where $n_b$ denotes the bulk molecular particle density of water and
$\bar{\alpha}(\omega)=\sum_j F_j/(E^2_j-\omega^2)$ denotes the mean
polarizability of an isolated water molecule, with $F_j$ and $E_j$
being the oscillator strength and excitation frequencies,
respectively, for excited state $j$. In the next section, we compute
$\bar{\alpha}(\omega)$ using time-dependent density-functional theory
and demonstrate that \eqref{eps_bulk}, which neglects the contribution
from the permanent dipole moments, indeed reliably describes the bulk
screening response of water to low-lying electronic excitations.  To
obtain excitation energies of the solute, we analyze the linear
response of \eqref{Sapprox} resulting in
\begin{equation}
\chi^{-1}=\chi^{-1}_{KS} - [\tilde{K} + f_{XC}],
\label{eq:newresponse}
\end{equation}
which lacks the subtleties appearing in Section~\ref{sec:implicit}
because the present model lacks any explicit environment dependence on
the external potential. Our final working equation is obtained by
expressing \eqref{eq:newresponse} in transition-space
notation\cite{Casida}, where the fused index $\kappa=(k,j)$ denotes a
transition between two equilibrium Kohn-Sham orbitals $\psi_j(\bm{r})$
and $\psi_k(\bm{r})$. We arrive at a self-consistent eigenvalue
problem \cite{Casida} for excitation energies $E_j$ of the solute,
\begin{equation}
\sum_\nu \left[ \delta_{\kappa\nu}\Delta\epsilon^2_{\nu}+
4\sqrt{\Delta\epsilon_\kappa \Delta\epsilon_\nu}\tilde{M}_{\kappa\nu}(E_j)
\right] C^{(j)}_{\nu} =E^2_j C^{(j)}_\kappa,
\label{JointLinearResponse}
\end{equation}
where $\Delta\epsilon_{\kappa}=\epsilon_j-\epsilon_k$ with
$\epsilon_k$ denoting the equilibrium orbital energies and the
eigenvector $C^{(j)}_\kappa$ determines the oscillator strength of the
transition\cite{Casida}. The coupling matrix is given by
\begin{align}
&\tilde{M}_{\kappa\nu}(\omega)= \nonumber\\
&\int_\Omega d\bm{r} \int_\Omega d\bm{r}' \Phi^*_\kappa(\bm{r})
\left[ \tilde{K}^R(\bm{r},\bm{r}',\omega)+ f^R_{XC}(\bm{r},\bm{r}',\omega) \right]
\Phi_\nu(\bm{r}')
\end{align}
with $\Phi_\kappa(\bm{r})=\psi^*_k(\bm{r})\psi_j(\bm{r})$ and
$f^R_{XC}$ denotes the retarded exchange-correlation kernel. Note that
even for a frequency-independent exchange-correlation kernel, the
solvent response makes $\tilde{M}$ \emph{frequency-dependent}.
Equation~\eqref{JointLinearResponse} is solved iteratively: setting
$\tilde{K}^R(E_j) = K$ yields an initial estimate $E_j^{(1)}$ for the
excitation energy. Next, we solve \eqref{JointLinearResponse} using
$\tilde{K}^R(E^{(1)}_j)$ and iterate until self-consistency is achieved.

\section{Application to formaldehyde in aqueous solution}
\label{sec:formaldehyde}
As a test case, we study the lowest singlet excited state of a
formaldehyde molecule in aqueous solution.  A number of theoretical
approaches have been applied to study solvato-chromic shifts of
formaldehyde in water
\cite{Barone,Mikkelsen,MennucciCammi,Mikkelsen2,Kawashima,Fukunaga,Naka}.
However, the agreement with experimental findings has generally been
unsatisfactory.

In this section, we first compute the mean polarizability of an
isolated water molecule from time-dependent density-functional theory
and obtain the frequency-dependent bulk dielectric function of liquid
water using the Clausius-Mossotti equation. Next, we explore the
excitations of formaldehyde in the gas phase and in solution using the
joint time-dependent density-functional theory described in the last
section.

All calculations are carried out in a plane wave basis with a cutoff
of $40$~hartree. We use Kleinman-Bylander pseudopotentials
\cite{KleinmanBylander} and a cubic supercell of length $20$~bohr.
For the ground state calculations we employ the local density
approximation \cite{KohnHohenberg,KohnSham} and for the excitations
the adiabatic local density approximation \cite{Casida}.

\subsection{Dielectric function of liquid water}

To compute excitation energies of a formaldehyde molecule in aqueous
solution, we need the frequency-dependent dielectric response of
water. According to \eqref{eps_bulk}, this requires the mean
polarizability of an isolated water molecule. We first carry out
ground-state calculations and fully relax the electronic and ionic
structure.  Then, we employ time-dependent density-functional theory
to obtain excitation energies and oscillator strengths using all 4
occupied Kohn-Sham orbitals, plus an additional 220 unoccupied
orbitals. Table~\ref{WaterVac} shows our results for the three lowest
singlet excitation energies of an isolated water molecule, and
compares them to previous theoretical work \cite{Bernasconi} and also
to experiment \cite{Bernasconi}. The discrepancy between theory and
experiment is around $1$~eV or larger for all excited states. The poor
performance of time-dependent density-functional theory for the water
molecule can be traced to the Rydberg character of the excitations,
which cannot be described in the adiabatic local density approximation
due to the incorrect asymptotic behavior of the exchange-correlation
potential at large distances \cite{Bernasconi,CasidaSalahub}.

Despite these problems, the adiabatic local density approximation
gives good results for the static polarizability and for the
low-frequency dielectric constant of \emph{liquid} water:
Table~\ref{Polarizability} compares our results for these quantities
with previous calculations \cite{Hu} and also with experiment
\cite{Hu,WaterReview,vanAlsenoy}. We observe that the
Clausius-Mossotti formula \eqref{eps_bulk} describes the dielectric
response of liquid water very well in the frequency range
corresponding to low-lying electronic excitations.

\begin{table}
\begin{center}
\setlength{\doublerulesep}{1\doublerulesep}
\setlength{\tabcolsep}{5\tabcolsep}
\caption{Comparison
 of the lowest singlet excitation energies of an isolated
 water molecule with previous theoretical work by Bernasconi
 \cite{Bernasconi} and experiment \cite{Bernasconi}. All results are
 given in eV.\\}  
\begin{ruledtabular}
 \begin{tabular}{ c c c}
   This work & Ref.\cite{Bernasconi} & Expt. \cite{Bernasconi} \\
   \hline
           &         &       \\
   6.47 & 6.39 & 7.4 \\
   7.74 & 7.78  & 9.1 \\
   8.01 & 8.05  & 9.7
 \end{tabular}
\end{ruledtabular}
\label{WaterVac}
\end{center}
\end{table}

\begin{table}
\begin{center}
\setlength{\doublerulesep}{1\doublerulesep}
\setlength{\tabcolsep}{3\tabcolsep}
\caption{Comparison of our results for the static mean polarizability
  of an \emph{isolated} water molecule and for the optical dielectric
  constant $\epsilon_{\rm opt} \equiv \epsilon_b(\omega=1~eV)$ of
  \emph{liquid water} with previous theoretical work by Hu {\em et
    al.}\cite{Hu} and experiment \cite{Hu,WaterReview,vanAlsenoy}. \\}
\begin{ruledtabular}
 \begin{tabular}{c c c c c}
   & Units & This work & Ref.\cite{Hu} & Expt. \cite{Hu,WaterReview,vanAlsenoy} \\
   \hline
     &                    &            &             &           \\
  $\bar{\alpha}(\omega=0)$& bohr$^3$  & 10.50   & 10.52  & 9.6-9.9 \\
  $\epsilon_{\rm opt}$  &           &  1.83  & ---   & 1.78 \\
 \end{tabular}
\end{ruledtabular}
\label{Polarizability}
\end{center}
\end{table}

\subsection{Formaldehyde in the gas phase}

Next, we explore the excitations of formaldehyde in the gas phase.
Table~\ref{FormVac} compares our results for the three lowest singlet
excited states with a previous calculation by Bauernschmitt and
Ahlrich \cite{AhlrichForm}, who also employ the adiabatic local
density approximation, and also with experimental findings
\cite{AhlrichForm,Barone}. We observe that our excitation energy for
the lowest state is relatively close to the experimental value, while
the higher states deviate more than $1$~eV from experiment. Again, the
relatively large deviation for the higher excited states is due to the
incorrect long-distance behavior of the exchange-correlation potential
in the adiabatic local density approximation.

\begin{table}
\begin{center}
\setlength{\doublerulesep}{1\doublerulesep}
\setlength{\tabcolsep}{5\tabcolsep}
\caption{Comparison of the three lowest singlet excitation energies of
an isolated formaldehyde molecule with previous theoretical work by
Bauernschmitt {\em et al.}\cite{AhlrichForm} and experiment
\cite{AhlrichForm,Barone}. All energies are given in eV.\\}
\begin{ruledtabular}    
\begin{tabular}{c c c}
   This work & Ref.\cite{AhlrichForm} & Expt. \cite{AhlrichForm,Barone} \\
   \hline
            &        &     \\
   3.66 & 3.64 & 3.8-4.2  \\
   5.68 & 5.93 & 7.13 \\
   6.78 & 6.79 & 8.14 \\
 \end{tabular}
\end{ruledtabular}
\label{FormVac}
\end{center}
\end{table}

Comparison of Tables~\ref{FormVac} and \ref{WaterVac} shows that the
lowest excitation energy of formaldehyde is several eV smaller than
the corresponding value for water. Therefore, in our joint
time-dependent density-functional calculations we only evaluate the
dielectric function at frequencies smaller than its first pole.  In
this region, $\epsilon_b(\omega)$ is close to unity and can be
approximated by a constant. This is a common approximation in ``polarizable
continuum model'' approaches, which for the non-equilibrium response
employ a frequency-independent dielectric function derived from the
index of refraction of water \cite{Barone}.

However, this approximation breaks down for solutes with higher-energy
excited states that are comparable or larger than the lowest excited
state of water. In this case, the full frequency dependence of the
dielectric response \emph{must} be retained and a self-consistent
solution of \eqref{JointLinearResponse} is necessary.

\subsection{Formaldehyde in aqueous solution}
We now apply joint time-dependent density-functional theory to
calculate excitations of a solvated formaldehyde molecule.  We use the
static joint density-functional theory of Petrosyan {\em et al.}
\cite{Petrosyan1} to determine the equilibrium electronic structure
neglecting ionic relaxations induced by the aqueous environment, which
as was shown by Kongsted {\em et al.} only lead to shifts in the
excitation energies of about $0.01$~eV \cite{Mikkelsen}.
Table~\ref{Dipole} shows that the equilibrium dipole moment obtained
in our calculation is in excellent agreement with previous theoretical
work\cite{Kawashima}.

\begin{table}
\begin{center}
\setlength{\doublerulesep}{1\doublerulesep}
\setlength{\tabcolsep}{2\tabcolsep}
\caption{Comparison of our results for the equilibrium dipole moment
of formaldehyde in vacuum ($p_{\rm vac}$) and aqueous solution
($p_{\rm solv}$) with experiment \cite{Matsika} and previous
theoretical work \cite{Kawashima}. Dipole moments are given in
bohr.\\}
\begin{ruledtabular}
 \begin{tabular}{c c c c}
    & This work & Ref. \cite{Kawashima} & Expt. \cite{Matsika} \\
   \hline
                           &           &                  \\
  $p_{\rm vac} $  & 0.91   & 0.90 & 0.91\\
  $p_{\rm solv}$  & 1.32   & 1.32 & ---  \\
 \end{tabular}
\end{ruledtabular}
\label{Dipole}
\end{center}
\end{table}

We then solve the linear response equation\eqref{JointLinearResponse}
of joint time-dependent density-functional theory self-consistently,
as described in the last section. The lowest excitation energy is
converged to within $0.01$~eV after two iterations. Table~\ref{Shifts}
summarizes our results for the excitation energies in vacuum and
solution and the resulting solvato-chromic shift, which is in good
agreement with the experimental value\cite{Mikkelsen}.

\begin{table}
\begin{center}
  \setlength{\doublerulesep}{1\doublerulesep}
  \setlength{\tabcolsep}{3\tabcolsep}
  \caption{Lowest singlet excitation energy of formaldehyde in
    vacuum ($E_{\rm vac}$) and aqueous solution ($E_{\rm solv}$)
    obtained from joint time-dependent density-functional theory.
    The resulting solvato-chromic shift ($E_{\rm solv}-E_{\rm vac}$)
    is shown in the third column and compared to
    experiment\cite{Mikkelsen}. All results are given in eV.\\}
  \begin{ruledtabular}
    \begin{tabular}{c c c c}
      $E_{\rm vac}$ & $E_{\rm solv}$ & Shift & Expt. \cite{Mikkelsen}\\
      \hline
      &                  &         &         \\       
      3.66 & 3.83 &  0.17 & 0.21\\
    \end{tabular}
  \end{ruledtabular}
  \label{Shifts}
\end{center}
\end{table}

To physically understand the observed solvato-chromic blue-shift, we
express the excitation energy as
\begin{equation}
E_{\rm vac/solv}=\Delta \epsilon_{\rm vac/solv} + \gamma_{\rm vac/solv},
\end{equation}
where $\Delta \epsilon_{\rm vac/solv}$ denotes orbital energy
difference in vacuum or solution and $\gamma_{\rm vac/solv}$ denotes
the correction from (joint) time-dependent density-functional
theory. For gas-phase formaldehyde, $\Delta \epsilon_{\rm
vac}=3.34$~eV already gives a reasonable approximation to $E_{\rm
vac}=3.66$~eV with the correction $\gamma_{\rm vac}=0.32$~eV being
relatively small.  In solution, we find $\Delta \epsilon_{\rm
solv}=3.56$~eV and $\gamma_{\rm solv}=0.27$~eV. We conclude that the
solvato-chromic shift is determined mostly by the change of the
orbital energy differences, $\Delta \epsilon_{\rm vac}-\Delta
\epsilon_{\rm solv}=0.22$~eV which is quite close to the total
solvato-chromic shift of $0.17$~eV, while the correction term changes
relatively little (only 0.05~eV).

The change of $\Delta\epsilon$ upon solvation is caused by the
different coupling of ground and excited states to the aqueous
environment: the ground state has a large dipole moment ($p^{\rm
vac}_{\rm gs}=0.91$~bohr) and couples strongly to the aqueous
environment leading to a large negative solvation energy, while the
dipole moment of the excited state ($p^{\rm vac}_{\rm ex}=0.14$~bohr)
and the resulting solvation energy are much smaller.

Table~\ref{Compare} compares our result for the solvato-chromic shift
of the lowest singlet excitation with previous theoretical work
\cite{Barone,Mikkelsen2,MennucciCammi,Fukunaga,Kawashima,Naka} and
also with experiment \cite{Mikkelsen}. In contrast to our calculation,
which gives very good agreement with the experimental data,
calculations employing the ``polarizable continuum model'' to describe
the solvent underestimate the shift \cite{Barone,MennucciCammi}, while
approaches, which treat the solvent atomistically within a
supermolecular approach\cite{Fukunaga,Kawashima}, typically
overestimate the shift. We point out that both methods suffer from
weaknesses which are absent in our approach. In particular, the
results of the ``polarizable continuum model'' approach depend
sensitively on the chosen cavity size and shape \cite{Bader}. This
indicates that a more realistic description of the solvent response in
the vicinity of the solute is of great importance. Atomistic solvent
models, on the other hand, offer a reliable description of the solvent
structure close to the solute, but to compute converged thermodynamic
averages the sampling of many solvent configurations is required. In
addition, supermolecular approaches which model the solvent by a
finite cluster surrounding the solute do not capture the long-range
dielectric response of the solvent.

Our approach includes both long-range screening effects and a reliable
description of the solvent response close to the solute. Similarly,
Naka {\em et al.} \cite{Naka}, who employ the reference
interaction-site model to describe the solute, and Kongsted {\em et
  al.} \cite{Mikkelsen2}, who combine an atomistic treatment of the
first solvation shells with the ``polarizable continuum model,'' also
obtain solvato-chromic shifts in good agreement with
experiment. However, unlike our action functional, these models are
not derived from first principles: instead they start out with a
partioning of the action and then approximate each contribution
typically with a different level of theory making it difficult to
judge the limits of their applicability \emph{a priori} and to
systematically improve upon them. For example, Naka \emph{et al.}
combine a CASSCF treatment of the solute electrons with an
electrostatic coupling scheme between solute and solvent and the
reference-interaction site model for the solvent structure
\cite{Naka}.

\begin{table}
\begin{center}
\setlength{\doublerulesep}{1\doublerulesep}
\setlength{\tabcolsep}{3\tabcolsep}
\caption{Comparison of our joint time-dependent density-functional
theory results, previous theoretical work
\cite{Barone,MennucciCammi,Kawashima,Fukunaga,Naka,Mikkelsen2} and
experiment \cite{Mikkelsen} for the solvato-chromic shift of the
lowest singlet excited state of formaldehyde in aqueous
solution. The second column lists the solvent model: joint
time-dependent density-functional theory (JTDDFT), the ``polarizable
continuum model'' (PCM), the supermolecular approach (SM) or the
reference interaction-site model (RISM). All results are given in
eV.\\}
\begin{ruledtabular}
 \begin{tabular}{c c c}
    & Method & Shift \\
   \hline
                         &  &         \\
    Ref. \cite{Barone}   & PCM & 0.12 \\
    Ref. \cite{MennucciCammi} & PCM & 0.12 \\
    Ref. \cite{Kawashima} & SM & 0.33\\
    Ref.  \cite{Fukunaga} & SM &0.39 \\
    Ref. \cite{Naka} & RISM &0.25 \\   
    Ref.  \cite{Mikkelsen2}&SM+PCM & 0.23 \\   
    This work       & JTDDFT & 0.17 \\
    Expt. \cite{Mikkelsen}  & ---  & 0.21 \\
 \end{tabular}
\end{ruledtabular}
\label{Compare}
\end{center}
\end{table}
\section{Summary and conclusions}
\label{sec:summary}
In sum, we describe the construction of a {\em joint} time-dependent
density-functional theory for the modelling of solute-solvent
systems. We derive coarse-grained action functionals by eliminating
environment degrees of freedom. This procedure enables us to examine
the underlying assumptions and uncover previously ignored functional
dependencies in popular approaches such as the time-dependent
``polarizable continuum model'' and to explore their domains of
validity. In particular, we find additional contributions to the
action functional which are typically neglected in standard
approaches. Also, in order to replace the full frequency dependent
solvent response by a high-frequency dielectric \emph{constant}, as is
often done in standard approaches, the excitation energy of the solute
has to be far from the poles of the solvent dielectric
function. Otherwise, a self-consistent solution of the linear response
equation is necessary.

We also introduce an explicit, approximate action functional for the
modelling of electronic systems in aqueous solution. Application of
this functional to solvated formaldehyde leads to good agreement with
experiment for the solvato-chromic shift of the lowest singlet excited
state.  The novel implicit functional we introduced can now be applied
to more complicated systems, such as dyes in Gr\"atzel cells
\cite{Graetzel} or solvated DNA molecules \cite{Rubio,Improta}.

The framework is now in place for future work to develop approximate
forms for the explicit solvent functionals by making use of existing
forms of time-dependent functionals for classical liquids
\cite{Archer} and to generalize time-independent functionals for
molecular liquids developed by us \cite{LischnerArias1,LischnerArias2}
to include time-dependent response of the solvent electrons. The
resulting theory would then allow us to describe the nuclear dynamics
of the environment during the excitation and to compute solvation
relaxation functions which have been measured experimentally
\cite{Maroncelli1,Maroncelli2}.

\begin{acknowledgements}
J.L. acknowledges valuable discussions with N. W. Ashcroft who
suggested the use of the Clausius-Mossotti formula for the dielectric
constant. J.L. acknowledges financial support by DOE \#
DE-FG02-07ER46432.
\end{acknowledgements}
\bibliography{jtddft}
\end{document}